\documentstyle[12pt,epsf]{article}
\def\slashchar#1{\setbox0=\hbox{$#1$}           % set a box for #1
   \dimen0=\wd0                                 % and get its size
   \setbox1=\hbox{/} \dimen1=\wd1               % get size of /
   \ifdim\dimen0>\dimen1                        % #1 is bigger
      \rlap{\hbox to \dimen0{\hfil/\hfil}}      % so center / in box
      #1                                        % and print #1
   \else                                        % / is bigger
      \rlap{\hbox to \dimen1{\hfil$#1$\hfil}}   % so center #1
      /                                         % and print /
   \fi}                                         %
\catcode`@=11
\long\def\@caption#1[#2]#3{\par\addcontentsline{\csname
  ext@#1\endcsname}{#1}{\protect\numberline{\csname
  the#1\endcsname}{\ignorespaces #2}}\begingroup
    \small
    \@parboxrestore
    \@makecaption{\csname fnum@#1\endcsname}{\ignorespaces #3}\par
  \endgroup}
\catcode`@=12

\def\jfig#1#2#3{
 \begin{figure}
 \centering
 \epsfysize=4.6in
 \epsffile{#2}
 \caption{#3}
 \label{#1}
 \end{figure}}
\begin{document}

\baselineskip=18pt

\begin{titlepage}
\begin{flushright}
IZTECH-P/2006-02
\end{flushright}

\begin{center}
\vspace{1cm}

{\Large \bf Generalized Modified Gravity in Large Extra Dimensions
}

\vspace{0.5cm}

{\bf {\"O}nder Aslan and Durmu{\c s} A. Demir}

\vspace{.8cm}

{\it Department of Physics, Izmir Institute of Technology, IZTECH, TR35430, Izmir, Turkey}

\end{center}
\vspace{1cm}

\begin{abstract}
\medskip
We discuss effective interactions among brane matter induced by
modifications of higher dimensional Einstein gravity through the
replacement of Einstein-Hilbert term  with a generic function
$f\left({\cal{R}},{\cal{R}}_{A B} {\cal{R}}^{A B} , {\cal{R}}_{A B
C D} {\cal{R}}^{A B C D}\right)$ of the curvature tensors. We
determine gravi-particle spectrum of the theory, and perform a
comparative analysis of its predictions with those of the Einstein
gravity within Arkani-Hamed--Dvali--Dimopoulos (ADD) setup. We
find that this general higher-curvature quantum gravity theory
contributes to scatterings among both massive and massless brane
matter (in contrast to much simpler generalization of the Einstein
gravity, $f\left({\cal{R}}\right)$, which influences only the
massive matter), and therefore, can be probed via various
scattering processes at present and future colliders and directly
confronted with the ADD expectations. In addition to collision
processes which proceed with tree-level gravi-particle exchange,
effective interactions among brane matter are found to exhibit a
strong sensitivity to higher-curvature gravity via the
gravi-particle loops. Furthermore, particle collisions with
missing energy in their final states are found to be sensitive to
additional gravi-particles not found in Einstein gravity. In
general, road to a correct description of quantum gravity above
Fermi energies depends crucially on if collider and other search
methods end up with a negative or positive answer for the presence
of higher-curvature gravitational interactions.

\end{abstract}

\bigskip
\bigskip

\begin{flushleft}
IZTECH-P/2006-02 \\
January 2006
\end{flushleft}

\end{titlepage}

%%%%%%%%%%%%%%%%%%%%%%%%%%%%%%%%%%%%%%%%%%%%%%%%%%%%%%%%%%%%%%%
%\tableofcontents

\section{Introduction}
The extra spatial dimensions (large \cite{large-extra}, warped
\cite{warped-extra} or hyperbolic \cite{hyperbolic-extra}) have
proven useful in solving the gauge hierarchy problem within the
quantum gravitational framework. In particular, large extra
dimensions induce Newton's constant in four dimensions from ${\rm
TeV}$ scale Einstein gravity via the large volume of the extra
space. The basic setup of this scenario $i.e.$
Arkani-Hamed--Dimopoulos--Dvali (ADD) scenario \cite{large-extra},
is that $(1+3)$--dimensional universe we live in is a
field-theoretic brane \cite{rubakov} which traps all flavors of
matter except the SM singlets $e.g.$ the  graviton and
right-handed neutrinos. As long as the surface tension of the
brane does not exceed the fundamental scale $\overline{M}_D$ of
$D$--dimensional gravity, at distances $\gg 1/\overline{M}_D$ the
spacetime metric $g_{A B}$ remains essentially flat. In other
words, for singlet emissions (from brane) with transverse (to
brane) momenta $\left|\vec{p}_T\right| \ll \overline{M}_D$ the
background spacetime is basically Minkowski. Therefore, it is
admissible to expand $D$--dimensional metric about the flat
background
\begin{eqnarray}
\label{metric} g_{A B} = \eta_{A B} + 2 \overline{M}_D^{1-D/2} h_{A B}
\end{eqnarray}
where $\eta_{A B} =\mbox{diag.}\left(1, -1, -1, \cdots, -1\right)$
and $h_{A B}$ are perturbations. The gravitational sector is
described by the Einstein-Hilbert action
\begin{eqnarray}
\label{ah} S_{ADD}=\int d^{D}x\, \sqrt{-g} \left\{ - \frac{1}{2}
\overline{M}_D^{D-2} {\cal{R}}\left(g_{A B}\right) +
{\cal{L}}_{matter}\left(g_{A B}, \psi\right)\right\}
\end{eqnarray}
where $\psi$ collectively denotes matter fields localized on the
brane. There are various ways \cite{large-extra} to see that the
Planck scale seen on the brane is related to the fundamental scale
of gravity in higher dimensions via
\begin{eqnarray}
\label{planck} \overline{M}_{Pl} = \sqrt{V_{\delta}} \overline{M}_D^{1+\delta/2}
\end{eqnarray}
which equals $(2 \pi R)^{1/2} \overline{M}_D^{1+\delta/2}$ when
$\delta\equiv D-4$ extra spatial dimensions are compactified over
a torus of radius $R$. Obviously, larger the $R$ closer the
$\overline{M}_D$ to the electroweak scale \cite{large-extra}. Upon
compactification, the higher dimensional graviton gives rise to a
tower of massive S, P and D states on the brane, and they
participate in various scattering processes involving radiative
corrections to SM parameters, missing energy signals as well as
graviton exchange processes. The collider signatures of these
processes have been discussed in detail in seminal papers
\cite{giudice,han}.

The ADD mechanism is based on higher dimensional Einstein gravity
with the metric (\ref{metric}). However, given general covariance
alone, there is no symmetry reason to guarantee that the action
density in (\ref{ah}) is unique. Indeed, general covariance does
not forbid the action density in (\ref{ah}) to be generalized to a
generic function $f\Big({\cal{R}},$ $\Box {\cal{R}},$ $\nabla_{A}
{\cal{R}} \nabla^{A} {\cal{R}},$ ${\cal{R}}_{A B} {\cal{R}}^{A
B},$ ${\cal{R}}_{A B C D} {\cal{R}}^{A B C D}, \dots\Big)$ of
curvature invariants. In fact, such modifications of Einstein
gravity have already been proposed and utilized for purposes of
improving the renormalizability of the theory
\cite{renorm,renorm2} and for explaining recent acceleration of
the universe \cite{cosmo,cosmox}.  Of course, once we depart from
the minimal Einstein-Hilbert regime there is no rule whatsoever
which can limit numbers and types of the invariants. Our approach
here, however, is to consider only those invariants which are of
lowest mass dimension and are quadratic contractions of the
curvature tensors: ${\cal{R}}_{A B} {\cal{R}}^{A B}$ and
${\cal{R}}_{A B C D} {\cal{R}}^{A B C D}$ in spite of fact that we
do not have any symmetry reason for not considering the
higher-derivative ones $\Box {\cal{R}},$ $\nabla_{A} {\cal{R}}
\nabla^{A} {\cal{R}},$ ${\nabla_C \cal{R}}_{A B} \nabla^C
{\cal{R}}^{A B},$ etc. In effect, we generalize Einstein-Hilbert
term to a generic function $f\left({\cal{R}}, {\cal{R}}_{A B}
{\cal{R}}^{A B}, {\cal{R}}_{A B C D} {\cal{R}}^{A B C D}\right)$
of the curvature invariants, and derive and analyze effective
interactions among brane matter induced by such modifications of
higher dimensional Einstein gravity.

The simplest generalization of (\ref{ah}) would be to consider a
generic function $f({\cal{R}})$ of the curvature scalar. This
possibility has been analyzed in detail in the recent work
\cite{fr}, and it has been found that $f({\cal{R}})$ gravity
effects are particularly pronounced and become distinguishable
from those of the Einstein gravity in scattering processes
involving massive brane matter $i.e.$ heavy fermions, weak bosons
and the Higgs boson (for recent work on Lovelock gravity see
\cite{rizzo}). The reason is that $f({\cal{R}})$ theory is
equivalent to Einstein gravity plus an independent scalar field
theory, and it is the propagation of this additional scalar that
causes observable differences between the $f({\cal{R}})$ gravity
and ADD setup in high energy processes \cite{fr}.

Here it is worth emphasizing that considering $f\left({\cal{R}},
{\cal{R}}_{A B} {\cal{R}}^{A B}, {\cal{R}}_{A B C D} {\cal{R}}^{A
B C D}\right)$ theory instead of $f\left({\cal{R}}\right)$ gravity
is not a straightforward generalization. The reason is that the
former is a four-derivative theory, and it is generically endowed
with a spin=2 ghost \cite{renorm2}. The presence of such
negative-norm states constitutes the main difference between the
two types of modified gravity theories, and goal of the present
work is to examine their signatures in high-energy processes in a
comparative fashion. Such ghosty states are obviously dangerous in
four dimensions (care should be payed to nonlinearities though)
especially at large distances \cite{cosmo,cosmox}; however, in a
higher dimensional setting, it is the experiment (at the LHC or
ILC) which will eventually establish presence or absence of such
states whereby providing a deeper understanding of yet-to-be found
quantum theory of gravity.

In this work we will consider a general modification of the
Einstein gravity, and discuss its physics implications in
comparison with the ADD and $f({\cal{R}})$ gravity setups. In Sec.
2 below we derive graviton propagator and describe how it
interacts with brane matter. Here we put special emphasis on
virtual graviton exchange. In Sec. 3 we study a number of higher
dimensional operators which are sensitive to modified gravity
effects. In Sec. 4 we briefly discuss some further signatures of
modified gravity concerning graviton production and decay as well
as certain loop observables on the brane. In Sec. 5 we conclude.

\section{Virtual Gravi-Particle Exchange}
The modification of the Einstein gravity we consider is
parameterized by
\begin{eqnarray}
\label{fr} S = \int d^{D}x\, \sqrt{-g} \left\{ - \frac{1}{2}
\overline{M}_D^{D-2} f\left({\cal{R}}, P, Q\right) +
{\cal{L}}_{matter}\left(g_{A B}, \psi\right)\right\}
\end{eqnarray}
where couplings to matter fields $\psi$ are identical to those in
(\ref{ah}). Here $P$ and $Q$, as in \cite{cosmo,cosmox}, stand,
respectively, for the quadratic contractions of the Ricci and
Riemann tensors:
\begin{eqnarray}
P = {\cal{R}}_{A B} {\cal{R}}^{A B}\; , \;\; Q = {\cal{R}}_{A B C
D} {\cal{R}}^{A B C D}
\end{eqnarray}
which contain four derivatives. The metric field obeys
\begin{eqnarray}
\label{eom} &&\left[ \nabla_A \nabla_B - g_{A B} \Box -
{\cal{R}}_{A B} \right] f_R\nonumber\\ &+& \left[2 \nabla_A
\nabla^C {\cal{R}}_{C B} - \Box {\cal{R}}_{A B}- g_{A B} \nabla^C
\nabla^D {\cal{R}}_{C D} - 2 {\cal{R}}_{C A}
{\cal{R}}^{C}_{B}\right] f_P\nonumber\\
&+& \left[ 4 \nabla^C \nabla^D {\cal{R}}_{C B A D} - 2
{\cal{R}}_{C D E A} {\cal{R}}^{C D E}_{B}\right] f_Q + \frac{1}{2}
f g_{A B}
 = \frac{ {\cal{T}}_{A B}}{\overline{M}_{D}^{D - 2}}
\end{eqnarray}
where $f_R \equiv \partial f/\partial R$, $f_P \equiv \partial
f/\partial P$, $f_Q \equiv \partial f/\partial Q$, and
\begin{eqnarray}
\label{tab} {\cal{T}}_{A B} = - \frac{2}{\sqrt{-g}} \frac{\delta \left(\sqrt{-g}
{\cal{L}}_{matter}\right)}{\delta g^{A B}} = \delta^{\delta}(\vec{y}) \delta_{A}^{\mu} \delta_{B}^{\nu} T_{\mu
\nu}(z)
\end{eqnarray}
is the stress tensor of the brane matter where $y_{i}$ and $z_{\mu}$ stand, respectively, for coordinates in
extra space and on the brane. The second equality here reflects the fact that entire energy and momentum are
localized on the brane. Clearly, energy-momentum flow has to be conserved $\nabla^{A} {\cal{T}}_{A B} =0$, and
this is guaranteed to happen provided that $\nabla^{\mu} T_{\mu \nu} = 0$.

Obviously, the equations of motion (\ref{eom}) reduce to Einstein
equations when $f({\cal{R}}, P, Q)= {\cal{R}}$. In general, for
analyzing dynamics of small oscillations about a background
geometry, $g_{A B} = g^0_{A B}$ with curvature scalar
${\cal{R}}_{0}$, $f({\cal{R}}, P, Q)$ must be regular at
${\cal{R}}= {\cal{R}}_{0}$. In particular, as suggested by
(\ref{eom}), $f({\cal{R}}, P, Q)$ must be regular at the origin
and $f(0, 0, 0)$ must vanish ($i.e.$ bulk cosmological constant
must vanish) for $f({\cal{R}}, P, Q)$ to admit a flat background
geometry.

For determining how higher curvature gravity (\ref{fr}) influences
interactions among the brane matter, it is necessary to determine
the propagating modes which couple to the matter stress tensor.
This requires expansion of the action density by using
(\ref{metric}) up to the desired order in $h_{A B}$. The zeroth
order term obviously vanishes. The terms first order in $h_{A B}$
vanish by the equations of motion (\ref{eom}). The quadratic part,
on the other hand, turns out to be
\begin{eqnarray}
\label{sh} S_{h} = \int d^{D} x \left[\frac{1}{2} h_{A B}(x) {\cal{O}}^{A B C D}(x) h_{C D}(x) -
\frac{1}{\overline{M}_{D}^{(D-2)/2}} h_{A B}(x) {\cal{T}}(x)^{A B}\right]
\end{eqnarray}
such that propagator of $h_{A B}(x)$, defined via the relation
\begin{eqnarray}
\label{prop-rel} {\cal{O}}_{A B C D}(x) {\cal{D}}^{C D E F}(x,
x^{\prime}) = \frac{1}{2} \delta^{D}(x-x^{\prime})
\left(\delta^{E}_{A} \delta^{F}_{B} + \delta^{E}_{B}
\delta^{F}_{A}\right)\,,
\end{eqnarray}
takes the form
\begin{eqnarray}
\label{prop} - i {\cal{D}}^{A B C D}(p^2) &=& {d_1(p^2)} \eta^{A
B} \eta^{C D}+{d_2(p^2)} \left(\eta^{A C} \eta^{B D} + \eta^{A D}
\eta^{B C}\right)\nonumber\\&+& {d_3(p^2)}\left(p^A p^B \eta^{C D}
+ \eta^{A B} p^C p^D\right)\nonumber\\&+&{d_4(p^2)}\left(\eta^{B
C} p^A p^D + \eta^{A D} p^B p^C + \eta^{A C} p^{B} p^D + \eta^{B
D} p^A p^C\right)\nonumber\\
&+&{d_5(p^2)} p^A p^B p^C p^D
\end{eqnarray}
where the form factors $d_{1,\dots,5}(p^2)$ depend on the
underlying theory of gravitation. In ADD setup, based on Einstein
gravity, they are given by $d_1(p^2)=-1/(D-2)p^2$, $d_2(p^2)=1/2
p^2$, $d_4(p^2)=(\xi-1)/2 p^4$, $d_{3}(p^2)=d_{5}(p^2) = 0$. In
$f\left({\cal{R}}\right)$ gravity none of them vanishes and their
explicit expressions can be found in \cite{fr}. In the framework
of modified gravity discussed here, they obtain nontrivial
structures, too. For graviton-mediated interactions among
brane-localized matter with conserved energy-momentum, only
$d_1(p^2)$ and $d_2(p^2)$ are relevant, and they are given by
\begin{eqnarray}
d_1(p^2) &=& -\frac{1}{(D-2) {f_R(0)} p^2} + \frac{1}{(D-1)
{f_R(0)} \left(p^2-m_1^2\right)} + \frac{1}{(D-1)(D-2) f_R(0) (p^2
- m_{\phi}^2)}\nonumber\\
d_2(p^2) &=& \frac{1}{2 f_R(0) p^2} - \frac{1}{2 f_R(0) (p^2 -
m_1^2)}
\end{eqnarray}
where we introduced the mass scales
\begin{eqnarray}
\label{mphi} m_0^2 = - \frac{4 f_{R}(0)}{f_{P}(0)-8 f_{R
R}(0)}\;,\;\; m_1^{2} = - \frac{4 f_R(0)}{f_P(0)+4
f_{Q}(0)}\;,\;\; m_{\phi}^2 = - \frac{(D-2) m_0^2 m_1^2}{(D-1)
m_1^2 + m_0^2}
\end{eqnarray}
parameterizing the non-minimal nature of the gravity theory
considered. The remaining form factors $d_{3,4,5}(p^2)$ can be
obtained from (\ref{prop-rel}) straightforwardly. Though it does
not appear in $d_1(p^2)$ and $d_2(p^2)$ above, in general, the
propagator depends on the gauge-fixing parameter $\xi$ following
from the gauge-fixing term
\begin{eqnarray}
{\cal{L}}_{g} = \frac{f_R(0)}{\xi} \eta^{A C}\left(\partial^{B}
h_{A B} - \frac{1}{2} \partial_{A} h_B^B\right) \left(\partial^{D}
h_{C D} - \frac{1}{2} \partial_{C} h_D^D\right)
\end{eqnarray}
added to the action density in  (\ref{sh}). Here, $f_R(0)$ is
introduced to match the terms generated by ${\cal{L}}_{g}$ with
the ones in (\ref{sh}). The de Donder gauge, $\xi =1$, is
frequently employed in quantum gravity.

Having determined the propagator, it is timely to analyze
gravi-particles in the system and their propagation
characteristics. The propagating modes and their properties are
determined by the pole structures of $d_1(p^2)$ and $d_{2}(p^2)$
(and by the remaining form factors $d_{3,4,5}(p^2)$ when the
longitudinal polarizations are taken into account). Indeed, the
pole at $p^2=0$ guarantees the existence of a massless $J=2$ mode
in $D$ dimensions. That this is the case directly follows from the
projector
\begin{eqnarray}
\label{proj1} \frac{1}{2}\left(\eta^{A C} \eta^{B D} + \eta^{A D}
\eta^{B C}\right) - \frac{1}{D-2}  \eta^{A B} \eta^{C D}
\end{eqnarray}
multiplying $1/p^2$. By restoring the longitudinal components via
the replacement $\eta_{A B} \rightarrow \eta_{A B} - p_A p_B /
p^2$ in each term of (\ref{proj1}) one ensures that the normal
mode under concern corresponds to a massless $J=2$ excitation, the
graviton \cite{renorm2}. In fact, when $f_R(0) = 1$, $f_{Q}(0)=0$,
$f_{RR}(0)=0$ and $f_{P}(0) = 0$ the whole propagator
(\ref{prop}), as it should, reduces to that computed in the
Einstein gravity \cite{giudice,han}.

The second terms of $d_1(p^2)$ and $d_2(p^2)$ combine to give a
massive $J=2$ propagating mode. Indeed, these two terms result in
weighing of $1/\left(p^2-m_1^2\right)$ with the projector
\begin{eqnarray}
\label{proj2} \frac{1}{2}\left(\eta^{A C} \eta^{B D} + \eta^{A D}
\eta^{B C}\right) - \frac{1}{D-1}  \eta^{A B} \eta^{C D}
\end{eqnarray}
which corresponds to a massive $J=2$ excitation in $D$ dimensions.
The most spectacular aspect of this propagator is that it has a
negative residue that is the excitation under concern is a ghost
represented by negative norm states in Hilbert space. This can be
cured by no choice of the model parameters because the sole and
obvious choice of negative $f_R(0)$ converts the massless graviton
discussed above into a ghost -- an absolutely unwanted situation
since then theory possesses no Einsteinian limit at any mass
scale. The existence of this tensorial ghost is a characteristic
property of higher curvature gravity consisting of Ricci and
Riemann tensors \cite{renorm2}, and it actually plays a rather
affirmative role in cancelling the divergences in loop
calculations in the same sense as the Pauli-Villars regulation
does in quantum field theory.

In addition to the aforementioned tensor modes, as evidenced by
the second line of $d_1(p^2)$, the particle spectrum also consists
of a scalar particle with mass-squared $m_{\phi}^2$. Indeed
$1/(p^2-m_{\phi}^2)$ is weighted by the projector
\begin{eqnarray}
\eta^{A B} \eta^{C D}
\end{eqnarray}
which guarantees the scalar nature of the propagating mode. This
mode is a tachyon as long as $m_0^2$ and $m_1^2$ have the same
sign otherwise it is a true scalar field. The parameter values
$i.e.$ signs of $m_1^2$ and $m_0^2$, competition between them as
well as various other factors give rise to several possibilities.
An interesting limit concerns $m_1^2 \rightarrow \pm \infty$,
which can be achieved by taking a special $f({\cal{R}}, P, Q)$
with $f_{P}(0) = - 4 f_{Q}(0)$ or $f_{P}(0) = 0=f_{Q}(0)$, then
the tensor ghost completely decouples from the spectrum. However,
the scalar field continues to accompany the tensor ghost with mass
$m_{\phi}^2 = - ((D-2/(D-1))m_0^2$ in agreement with \cite{fr}.
For generating the pure Einstein gravity one needs to send both
$m_0^2$ and $m_1^2$ to $\infty$ which necessitates $f_{P}(0),
f_{Q}(0), f_{RR}(0) \rightarrow 0$.

%%%%%%%%%%%%%%%%%%%%%%%%%%%%%%%%%%%%%%%%%%%%%
\jfig{boncuk_1}{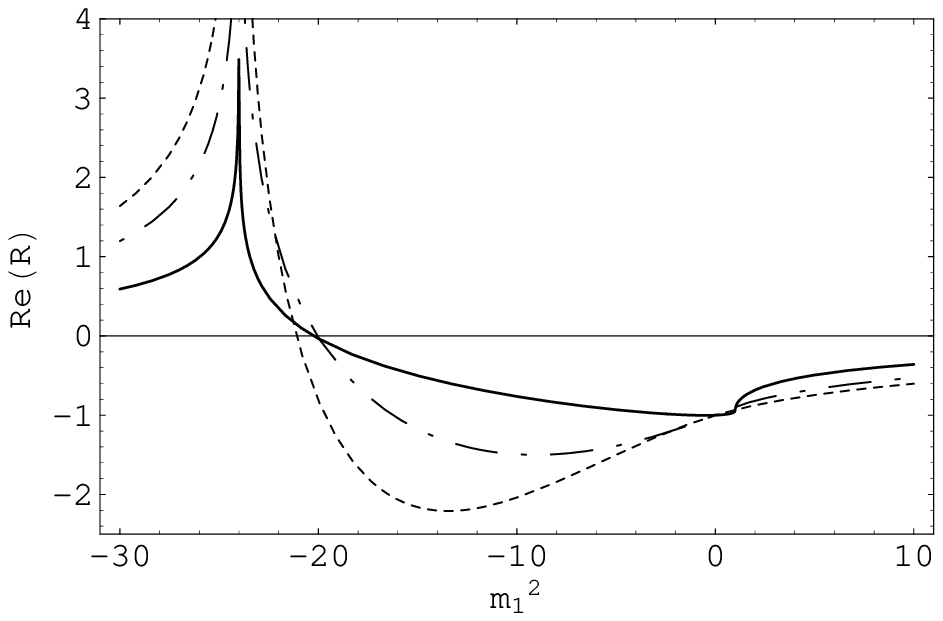}{ The dependence of
$\mbox{Re}\left[{{R}}(k^2)\right]$ on $m_{1}^2$ for $k^2= (1\,
{\rm TeV})^{2}$, $\Lambda = \overline{M}_{D} = 5\, {\rm TeV}$, and
$\delta=3$ (solid curve), $\delta=5$ (dot-dashed curve) and
$\delta=7$ (short-dashed curve). In the plot $m_{1}^2$ varies from
$- 30\, {\rm TeV}^2$ up to $+ 10\, {\rm TeV}^2$.}
%%%%%%%%%%%%%%%%%%%%%%%%%%%%%%%%%%%%%%%%%%%%%5

%%%%%%%%%%%%%%%%%%%%%%%%%%%%%%%%%%%%%%%%%%%%%
\jfig{boncuk2}{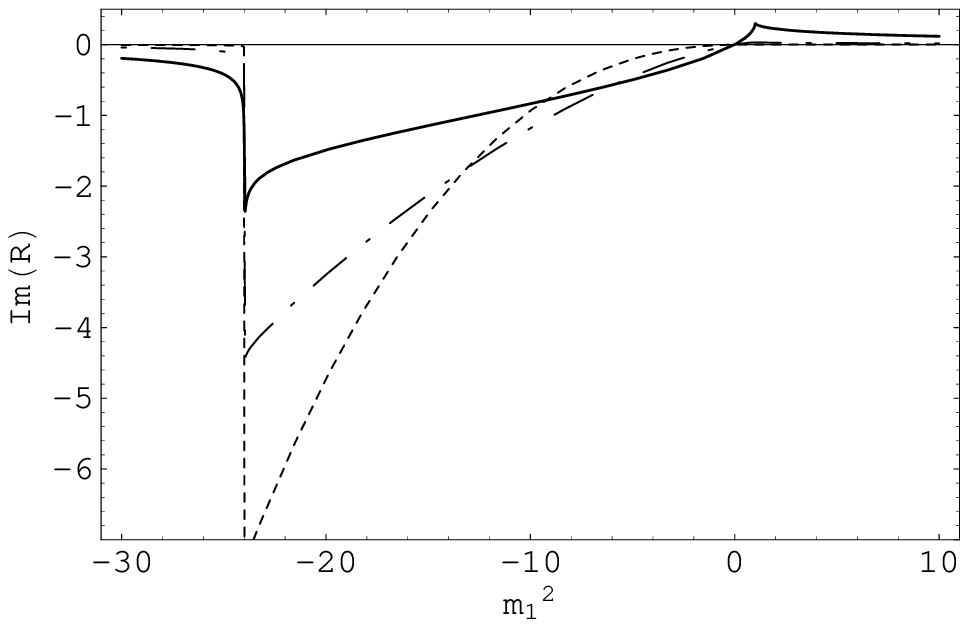}{ The same as in Fig. \ref{boncuk_1} but
for $\mbox{Im}\left[{{R}}(k^2)\right]$.}
%%%%%%%%%%%%%%%%%%%%%%%%%%%%%%%%%%%%%%%%%%%%%5

Having determined the gravi-particle spectrum of
$f\left({\cal{R}}, P, Q\right)$ gravity in $D$ dimensions, we
start analyzing the consequences of the compactness of the extra
space. Indeed, by letting extra space be torus-shaped with radius
$R$ as in the ADD mechanism, the matter stress tensor obeys the
Kaluza-Klein expansion
\begin{eqnarray}
{\cal{T}}_{A B}(x) = \sum_{n_1=-\infty}^{+\infty}\cdots \sum_{n_{\delta}=-\infty}^{+\infty} \int \frac{d^4
p}{(2\pi)^4} \frac{1}{\sqrt{V_{\delta}}} e^{- i \left(k\cdot z - \frac{\vec{n}\cdot\vec{y}}{R}\right)}
\delta_{A}^{\mu} \delta_{B}^{\nu} {T}_{\mu \nu}(k)
\end{eqnarray}
where $\left(n_1, \dots, n_{\delta}\right)$ is a $\delta$-tuple of
integers. Given this Fourier decomposition of the stress tensor,
the amplitude for an on-brane system $a$ to make a transition into
another on-brane system $b$ becomes
\begin{eqnarray}
\label{amp} {\cal{A}}(k^2) = \frac{1}{\overline{M}_{Pl}^2} \sum_{\vec{n}} T_{\mu \nu}^{(a)}(k)
{\cal{D}}^{\mu\nu\lambda\rho}\left(k^2 - \frac{\vec{n}\cdot\vec{n}}{R^2}\right) T_{\lambda \rho}^{(b)}(k)
\end{eqnarray}
where use has been made of (\ref{planck}) in obtaining
$1/\overline{M}_{Pl}^2$ factor in front. Though we are dealing
with a tree-level process the amplitude involves a summation over
all Kaluza-Klein levels due to the fact that these states are
inherently virtual because of their propagation off the brane.
Conservation of energy and momentum implies that only the first
two terms in the propagator (\ref{prop}) contributes to
(\ref{amp}), and  after performing summation the transition
amplitude takes the form
\begin{eqnarray}
\label{as} {\cal{A}}(k^2)&=& \frac{S_{\delta-1}}{(2
\pi)^{\delta}}\, \frac{1}{\overline{M}_D^4 f_{R}(0)}
\left(\frac{\Lambda}{\overline{M}_{D}}\right)^{\delta-2}
\Bigg\{\nonumber\\
&& {\cal{G}}\left(\frac{\Lambda}{\sqrt{k^2}}\right)
\left(T^{(a)}_{\mu \nu} T^{(b)\, \mu \nu} - \frac{1}{\delta +2}
T_{\mu}^{(a)\,\mu}
T_{\nu}^{(b)\,\nu}\right)\nonumber\\
&-& {\cal{G}}\left(\frac{\Lambda}{\sqrt{k^2-m_1^2}}\right)
\left(T^{(a)}_{\mu \nu} T^{(b)\, \mu \nu} - \frac{1}{\delta +3}
T_{\mu}^{(a)\,\mu}
T_{\nu}^{(b)\,\nu}\right)\nonumber\\
&+& \frac{1}{(\delta+2) (\delta+3)}
{\cal{G}}\left(\frac{\Lambda}{\sqrt{k^2-m_{\phi}^2}}\right)
T_{\mu}^{(a)\,\mu} T_{\nu}^{(b)\,\nu}\Bigg\}
\end{eqnarray}
which exhibits a huge overall enhancement
${\cal{O}}\left(\overline{M}_{Pl}^2/\overline{M}_D^2\right)$
compared to (\ref{amp}) due to the contributions of finely-spaced
Kaluze-Klein levels \cite{large-extra}. Here $S_{\delta-1}= (2
\pi^{\delta/2})/\Gamma(\delta/2)$ is the surface area of
$\delta$-dimensional unit sphere and $\Lambda$ (which is expected
to be ${\cal{O}}\left(\overline{M}_D\right)$ since above
$\overline{M}_D$ underlying quantum theory of gravity completes
the classical treatment pursued here) is the ultraviolet cutoff
needed to tame divergent summation over Kaluza-Klein levels. In
fact, ${\cal{A}}(q^2)$ exhibits a strong dependence on $\Lambda$,
as suggested by (see also the corresponding series expressions
derived in \cite{giudice,han})
\begin{eqnarray}
\label{time} {\cal{G}}\left(\frac{\Lambda}{\sqrt{q^2}}\right) =  -
i \frac{\pi}{2}
\left(\frac{q^2}{\Lambda^2}\right)^{\frac{\delta}{2}-1} +
\frac{\pi}{2}
\left(\frac{q^2}{\Lambda^2}\right)^{\frac{\delta}{2}-1} \cot
\frac{\pi \delta}{2} - \frac{1}{\delta-2}\,\, {}_2F_{1}\left(1,
1-\frac{\delta}{2}, 2 -
\frac{\delta}{2},\frac{q^2}{\Lambda^2}\right)
\end{eqnarray}
for $0\leq q^2 \leq \Lambda^2$, and
\begin{eqnarray}
\label{space}
 {\cal{G}}\left(\frac{\Lambda}{\sqrt{q^2}}\right) = \frac{1}{\delta} \frac{\Lambda^2}{q^2}\,\, {}_2F_{1}\left(1, \frac{\delta}{2}, 1
+ \frac{\delta}{2},\frac{\Lambda^2}{q^2}\right)
\end{eqnarray}
for  $q^2 <0$ or $q^2>\Lambda^2$. The imaginary part of
${\cal{G}}$, relevant for the timelike propagator (\ref{time}), is
generated by exchange of on-shell gravitons $i.e.$ those
Kaluza-Klein levels satisfying $q^2=\vec{n}\cdot\vec{n}/R^2$. On
the other hand, its real part follows from exchange of off-shell
gravitons. For spacelike propagator,  the scattering amplitude
(\ref{space}) is real since in this channel Kaluza-Klein levels
cannot come on shell.

%%%%%%%%%%%%%%%%%%%%%%%%%%%%%%%%%%%%%%%%%%%%%
\jfig{boncuk_3}{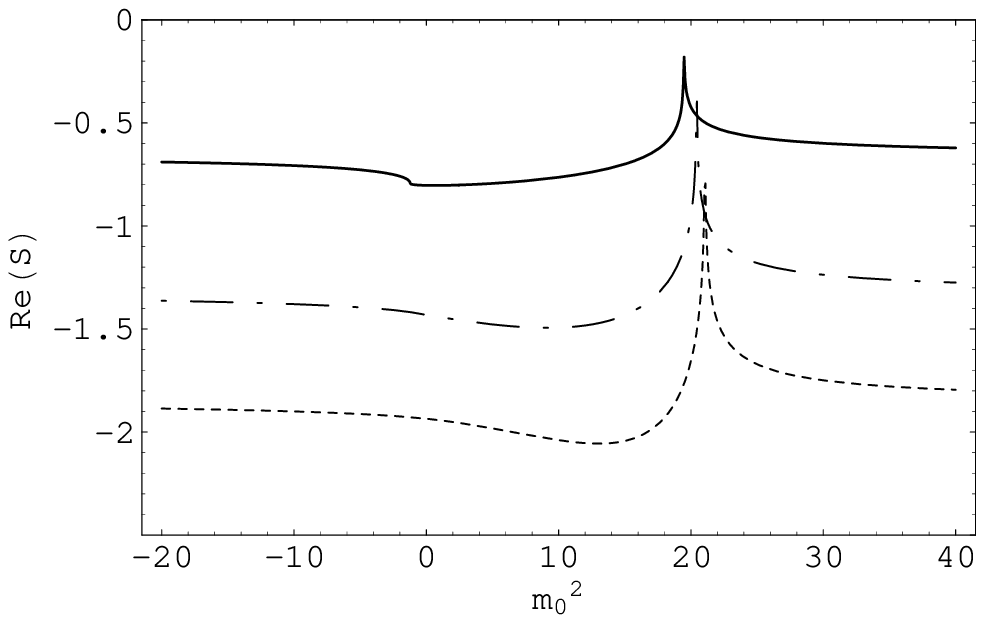}{ The dependence of
$\mbox{Re}\left[{{S}}(k^2)\right]$ on $m_{0}^2$ for $k^2= (1\,
{\rm TeV})^{2}$, $\Lambda = \overline{M}_{D} = 5\, {\rm TeV}$,
$m_1^2 = - 10\ {\rm TeV}^2$, and $\delta=3$ (solid curve),
$\delta=5$ (dot-dashed curve) and $\delta=7$ (short-dashed curve).
In the plot $m_{0}^2$ varies from $- 20\, {\rm TeV}^2$ up to $40\,
{\rm TeV}^2$, and $\mbox{Re}\left[{{S}}(k^2)\right]$ flattens for
large $|m_0^2|$.}
%%%%%%%%%%%%%%%%%%%%%%%%%%%%%%%%%%%%%%%%%%%%%

%%%%%%%%%%%%%%%%%%%%%%%%%%%%%%%%%%%%%%%%%%%%%
\jfig{boncuk4}{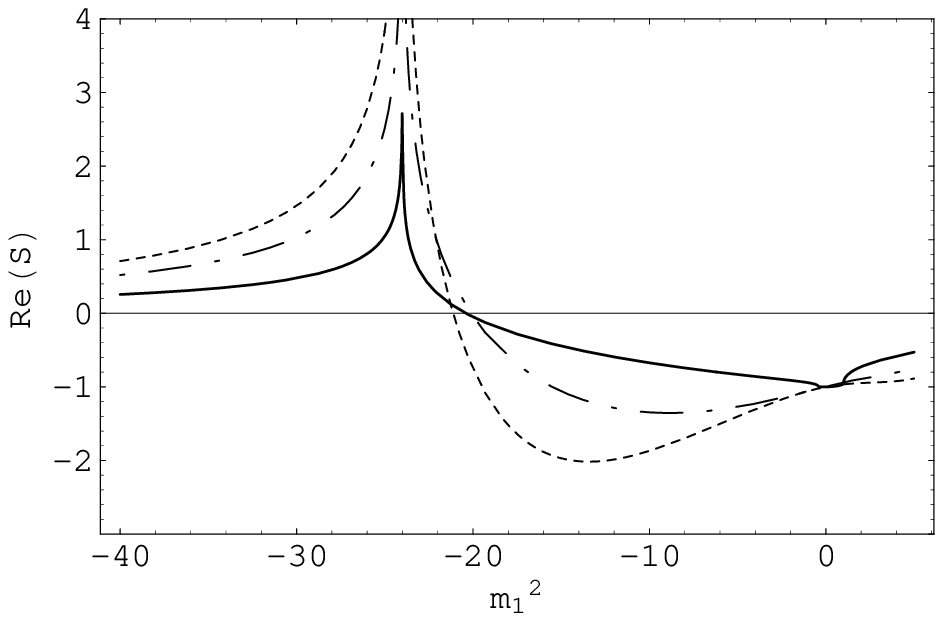}{ The dependence of
$\mbox{Re}\left[{{S}}(k^2)\right]$ on $m_{1}^2$ for $k^2= (1\,
{\rm TeV})^{2}$, $\Lambda = \overline{M}_{D} = 5\, {\rm TeV}$,
$m_0^2 = 5 m_1^2$, and $\delta=3$ (solid curve), $\delta=5$
(dot-dashed curve) and $\delta=7$ (short-dashed curve). In the
plot $m_{1}^2$ varies from $- 40\, {\rm TeV}^2$ up to $+ 5\, {\rm
TeV}^2$.}
%%%%%%%%%%%%%%%%%%%%%%%%%%%%%%%%%%%%%%%%%%%%%

The first line of ${\cal{A}}({k^2})$ in (\ref{as}), except for the
overall $1/f_{R}(0)$ factor in front, is identical to virtual
graviton exchange amplitude computed within the ADD setup. The
stress tensors of the on-brane systems $a$ and $b$ contribute to
the transition amplitude via their contractions $T^{(a)}_{\mu \nu}
T^{(b)\, \mu\nu}$ and via the multiplication of their traces
$T_{\mu}^{(a)\,\mu} T_{\nu}^{(b)\,\nu}$. While the former is
effective for any two systems of particles \cite{giudice,han}, the
latter can exist only for systems possessing conformal breaking
\cite{fr}.

The second line of (\ref{as}), induced by the exchange of a
massive graviton, is completely new in that it exists neither in
ADD \cite{giudice,han} nor in $f\left({\cal{R}}\right)$ gravity
setups. The presence of the operator $T^{(a)}_{\mu \nu} T^{(b)\,
\mu\nu}$ in this novel contribution proves particularly useful for
distinguishing this general modification of gravity from
$f\left({\cal{R}}\right)$ theory since the latter cannot induce
brane-localized operators which involve contractions of the stress
tensors.

The third  line of (\ref{as}) is generated by exchange of the
scalar graviton in the system. Its contribution always involves
traces of the stress tensors, and thus, for it to significantly
influence a scattering process conformal invariance must be broken
strongly (masses of the brane-localized fields must be a
significant fraction of $\overline{M}_D$), as has been analyzed in
detail elsewhere \cite{fr}.

It may be of practical use to illustrate how novel structures
induced by $f\left({\cal{R}}, P, Q\right)$ gravity compare with
the ones already present in the ADD setup. As has been emphasized
above, the higher-curvature gravity theory under concern modifies
the coefficients of both $T^{(a)}_{\mu \nu} T^{(b)\, \mu\nu}$ and
$T_{\mu}^{(a)\,\mu} T_{\nu}^{(b)\,\nu}$. The former is
particularly useful for collider searches  as well as effective
operators at low-energies since it does not require systems $a$
and $b$ to consist of massive brane matter. In this respect, it
could be useful to dwell on the coefficient of $T^{(a)}_{\mu \nu}
T^{(b)\, \mu\nu}$ for determining how $f\left({\cal{R}}, P,
Q\right)$ gravity contribution compares with the ADD prediction.
This we do by plotting the real and imaginary parts of
\begin{eqnarray}
\label{rfunc} {{R}}(k^2) = -
\frac{{\cal{G}}\left(\frac{\Lambda}{\sqrt{k^2-m_1^2}}\right)}{{\cal{G}}\left(\frac{\Lambda}{\sqrt{k^2}}\right)}
\end{eqnarray}
as a function of $m_1^2$ by taking, in accord with the future
collider searches, $k^2 = (1\, {\rm TeV})^{2}$ and $\Lambda =
\overline{M}_{D} = 5\, {\rm TeV}$. Their variations are plotted in
Figs. \ref{boncuk_1} and \ref{boncuk2} where $m_1^2$ is let vary
from $- 30\, {\rm TeV}^2$ up to $+ 10\, {\rm TeV}^2$ for each
number of extra dimensions considered: $\delta=3$ (solid),
$\delta=5$ (dot-dashed) and $\delta=7$ (short-dashed). These
figures make it clear that massive graviton (a ghosty tensor mode
special to $f\left({\cal{R}}, P, Q\right)$ gravity) exchange
significantly dominates, if not competes, the massless graviton
(the only propagating mode in ADD setup) exchange when the former
is a tachyon with mass-squared $\sim - 0.5 \Lambda^2$ ( excluding
the rather narrow peak at $m_{1}^2 = - 24\, {\rm TeV}^{2}$ which
corresponds to resonating of the transition amplitude by
Kaluza-Klein levels with mass-squared $=k^2-m_{1}^2=\Lambda^2$).
The ghosty nature of the massive graviton affects only the sign of
(\ref{rfunc}) whereas its tachyonic nature gives rise to a
spectacular enhancement in $R(k^2)$ which in turn enables one to
disentangle $f\left({\cal{R}}, P, Q\right)$ gravity effects from
those of the Einstein gravity in high-energy collider environment.
From (\ref{mphi}) it is clear that a negative $m_1^2$ implies a
positive $f_P(0) + 4 f_{Q}(0)$ since $f_R(0)$ must be positive for
preventing massless graviton from becoming a ghost.

We now turn to discussion of the coefficient of
$T_{\mu}^{(a)\,\mu} T_{\nu}^{(b)\,\nu}$ in (\ref{as}) for
determining impact of $f\left({\cal{R}}, P, Q\right)$ gravity
relative to Einstein gravity. We quantify analysis by examining
the ratio
\begin{eqnarray}
\label{sfunc} S(k^2) = \frac{\mbox{Coefficient of}\;
T_{\mu}^{(a)\,\mu} T_{\nu}^{(b)\,\nu}\; \mbox{from 2nd and 3rd
lines of}\; (\ref{as})}{\mbox{Coefficient of}\; T_{\mu}^{(a)\,\mu}
T_{\nu}^{(b)\,\nu}\; \mbox{from 1st line of}\; (\ref{as})}
\end{eqnarray}
in a way similar to (\ref{rfunc}). This quantity does not have a
direct meaning in interpreting the scattering rates of massive
brane matter as they receive contributions from $T^{(a)}_{\mu \nu}
T^{(b)\, \mu\nu}$, too. Nevertheless, for determining effects of
higher-curvature gravity it could be instructive to determine how
$S(k^2)$ depends on various model parameters. Depicted in Fig.
\ref{boncuk_3} is the variation of $\mbox{Re}\left[S(k^2)\right]$
with $m_0^2$ for $m_1^2 = -10 {\rm TeV}^2$, $k^2 = (1\, {\rm
TeV})^{2}$ and $\Lambda = \overline{M}_{D} = 5\, {\rm TeV}$  for
each number of extra dimensions considered: $\delta=3$ (solid),
$\delta=5$ (dot-dashed) and $\delta=7$ (short-dashed). The figure
shows it manifestly that $f\left({\cal{R}}, P, Q\right)$ gravity
contributions completely dominate the one found in the ADD setup
for large negative $m_0^2$. The extra gravi-scalar, not found in
Einstein gravity, results in an enhancement in scattering
amplitudes of massive brane matter.

We also plot $m_1^2$ dependence of $S(k^2)$ in Fig. \ref{boncuk4}
by taking  $m_0^2 = 5 m_1^2$ and keeping other parameters as in
Fig. \ref{boncuk_3}. Here, unlike the case study depicted in Fig.
\ref{boncuk_3}, gravi-particles decouple from the spectrum at
large $m_1^2$ due to the fact that $m_0^2$ varies in proportion
with $m_1^2$. The figure suggests that $f\left({\cal{R}}, P,
Q\right)$ gravity contribution is particularly enhanced in
negative $m_1^2$ domain especially when $m_1^2 \sim - 25\ {\rm
TeV}^2$. From the figures and their accompanying discussions one
therefore concludes that,  higher-curvature gravity theory
(\ref{fr}) provides additional gravi-particles and they  result in
significant enhancements in virtual gravi-particle exchange
amplitudes with respect to both Einstein \cite{giudice,han} and
$f\left({\cal{R}}\right)$ gravity theories.

In the next section we will survey and briefly discuss certain
observables (concerning especially collider searches for extra
dimensions) in light of the virtual gravi-particle exchange
amplitude  (\ref{as}) and its discussions and illustrations via
the figures.

\section{Effects of Gravi-Particle Exchange on Brane Processes}
It might be instructive to discuss in some length certain higher
dimensional operator structures which can leave significant impact
on scatterings among brane-localized matter. From (\ref{as}) it is
clear that virtual gravi-particle exchange between two systems of
brane matter leads to dimension-8 operators $T^{(a)}_{\mu \nu}
T^{(b)\, \mu\nu}$ and $T_{\mu}^{(a)\,\mu} T_{\nu}^{(b)\,\nu}$.
These additional interactions can open up novel scattering
channels not found in the SM or modify the existing ones in an
observable way. Therefore, they are of potential importance for
collider as well as precision physics of modified gravitational
interactions in higher dimensions.

It may be convenient to group and analyze effects of higher
dimensional operators according to the virtualities of the
gravi-particles involved, as we do in the following subsections.

\subsection{Tree-level effects of Virtual Gravi-Particles:}
The tree-level gravi-particle exchange, as has been detailed in
the last section, gives rise to anomalous interactions among brane
matter species \cite{giudice,han,fr}. The on-brane processes are
entirely tree-level ones in such processes; however, they exhibit
appreciable sensitivity to gravi-particle exchange due to rather
high virtualities that gravi-particles obtain via their
propagation through the extra dimensions. In general, tree-level
gravi-particle exchange induces various modifications in
scattering processes, and they may be detected at colliders or in
other experiments \cite{review}. At an $e^+ e^-$ collider, for
instance, pair-productions of gauge bosons ($e.g.$ $e^+ e^-
\rightarrow V V$ where $V=\gamma, Z, W$) and of fermions ($e.g.$
$e^+ e^- \rightarrow t \overline{t}$ or any other quark or lepton)
prove particularly useful for disentangling gravi-particle
effects. In fact, existing results from LEP experiments already
provide precise bounds on such effects from pair-productions of
gauge bosons and fermions \cite{lep}. There also exist promising
scattering processes at hadron ($p p$ collisions at the LHC and $p
\overline{p}$ collisions at Tevatron \cite{tevat}) and
lepton-hadron ($e p$ collisions at HERA \cite{hera}) colliders
which can probe gravi-particle effects at different energy scales
with different particle species. In addition, there exist various
phenomena, ranging from rare decays to supernovae and to ultra
high energy cosmic rays, by which one can put bounds of varying
strength on extra dimensions and nature of the gravitational
theory in higher dimensional bulk.

It may be useful to examine some generic scattering processes for
determining their power of disentangling the $f\left({\cal{R}}, P,
Q\right)$ gravity effects. For instance, generic scattering
amplitude ${\cal{A}}\left(\psi_{a}(k_1) \overline{\psi}_a (k_2)
\rightarrow \psi_{b}(q_1) \overline{\psi}_b (q_2)\right)$ for two
identical fermions can be directly obtained via the replacements
\begin{eqnarray}
\label{replace-fermion} T^{(a)}_{\mu \nu} T^{(b)\, \mu \nu}
&\rightarrow& \frac{1}{8} \Bigg[ (k_1+k_2)\cdot (q_1+q_2)
\overline{\psi}_a(k_2) \gamma^{\mu} \psi_a(k_1)\;
\overline{\psi}_b(q_2) \gamma_{\mu} \psi_b(q_1) \nonumber\\&+&
\overline{\psi}_a(k_2) ( \slashchar{q}_1+
\slashchar{q}_2)\psi_a(k_1)\; \overline{\psi}_b(q_2) (
\slashchar{k}_1+
\slashchar{k}_2)\psi_b(q_1)\Bigg] \nonumber\\
T_{\mu}^{(a)\,\mu} T_{\nu}^{(b)\,\nu} &\rightarrow& m_{\psi_a}
m_{\psi_b} \overline{\psi}_a(k_2) \psi_a(k_1)\;
\overline{\psi}_b(q_2) \psi_b(q_1)
\end{eqnarray}
in the tree-level gravi-particle exchange amplitude (\ref{as}).
These replacements correspond to $s$-channel gravi-particle
exchange with $k=k_1-k_2=q_2-q_1$, and depending on the quantum
numbers of $\psi_a$ and $\psi_b$ it could be necessary to include
$t$ and $u$ channel contributions, too. Its comparison with the
corresponding amplitude in ADD setup \cite{giudice,han} reveals
that fermion-fermion scattering now proceeds with additional
structures provided by second and third  lines of (\ref{as}). They
can compete in size with the ADD amplitude for certain parameter
values, especially when the scattering energy $\sqrt{k^2}$
compares with the new gravitational scales $m_0$ or $m_1$
\cite{fr}. A highly interesting aspect of (\ref{as}) with the
replacements (\ref{replace-fermion}) is that effects of the
modified gravity (due to the massive ghosty $J=2$ graviton)
survive even in the limit of massless fermions. This is , as one
recalls form \cite{fr}, not the case for $f\left({\cal{R}}\right)$
gravity to which only scatterings of the massive brane matter
exhibit sensitivity. This property of $f\left({\cal{R}}, P,
Q\right)$ gravity is important in that its effects can be directly
probed at high-energy colliders (where colliding beams of matter
are essentially massless) and effective higher-dimensional
operators consisting of light fermions. Indeed, LEP-favored modes
$e^+ e^- \rightarrow f \overline{f}$ ($f = e, \mu, b, t, \cdots$)
or Drell-Yan annihilation of quarks at hadron colliders are golden
modes for detecting modified gravity effects thanks to the fact
that fermions (of systems $a$ or $b$) under concern are massless.
In general, independent of if the brane matter is massive or
massles, there are certain parameter values for which
$f\left({\cal{R}}, P, Q\right)$ gravity contributions get
significantly enhanced  and thus become more easily observable
with respect to Einstein gravity effects as can be seen from Figs.
\ref{boncuk_1}, \ref{boncuk2}, \ref{boncuk_3} and \ref{boncuk4}.

The $2 \rightarrow 2$ fermion scattering example above can be
generalized to any other SM particle, and the relative
enhancements/suppressions in their rates are always governed by
(\ref{as}) with supplementary illustrations given in the figures.
A dedicated search for extra dimensions via virtual gravi-particle
exchange processes requires a global analysis of various collider
processes \cite{lep,tevat,hera}. The most advantageous aspect of
$f\left({\cal{R}}, P, Q\right)$ gravity is the separability of
massive ghosty graviton contribution from those of the remaining
gravi-particles via the measurement of the scattering rates of
massless (or more precisely much lighter than the fundamental
scale of gravity)  matter species.

\subsection{Loop-level effects of Virtual Gravi-Particles:} The
loop-level processes on the brane proceed with looping brane
matter and/or gravi-particles where the latter are now virtual
both in ordinary and extra dimensions. These effects can give rise
to  corrections to the existing SM amplitudes as in, for instance,
electroweak precision observables (particle self energies,
interaction vertices, box diagrams) and rare decays
\cite{giudice,han,fr}. In fact, the dedicated analysis of
\cite{giudicex} shows that gravi-particle loop effects can become
more important than their tree-level effects since they can induce
potentially important dimension-6 operators with double
gravi-particle exchange. This lower-dimension operator can arise
in fermion, gauge boson as well as Higgs sectors of the SM.

For illustrating the impact of $f\left({\cal{R}}, P, Q\right)$
gravity, consider dimension-6 four-fermion operator $(1/2)
\overline{f} \gamma_{\mu} \gamma_5 f\; \overline{f^{\prime}}
\gamma^{\mu} \gamma_5 f^{\prime}$ ($f$, $f^{\prime}$ standing for
light quarks or leptons) which has been shown to follow from
double gravi-particle exchange in \cite{giudicex}. A direct
calculation shows it is quite sensitive to exchange novel
propagating degrees of freedom in higher curvature gravity.
Indeed, for massless fermions, for instance, coefficient of this
operator for $f_{R}(0)=1$ is $1.9$, $2.6$, and $2.7$ times larger
than the coefficient of the same operator computed in the ADD
setup \cite{giudicex} for $m_1^2 = - 14\ {\rm TeV}^2$, $\Lambda =
5\ {\rm TeV}$ and $\delta= 3, 5, $ and $7$ extra dimensions. The
reason for this, as illustrated in Figs. \ref{boncuk_1} and
\ref{boncuk2}, is the enhancement of $f\left({\cal{R}}, P,
Q\right)$ gravity contributions compared to Einstein gravity at
this specific value of $m_1^2$. Clearly, existing experimental
results on contact interactions, dijet and dilepton production
processes as well as lepton-hadron scattering rates can put
stringent limits on the model parameters $\delta$, $\Lambda$,
$\overline{M_D}$, $m_1^2$ and $f_R(0)$. The analysis of
\cite{giudicex} shows that the strongest bounds come from LEP
results on contact interactions \cite{lep2x}.

Repeating, at the loop-level gravi-particle exchange gives rise to
observable modifications on various phenomena testable at the
present and future collider studies. Therefore, essentially what
remains to be done is to perform a global analysis of the
observables so as to achieve bounds or exclusion limits on
$f\left({\cal{R}}, P, Q\right)$ gravity parameters.

\subsection{Effects of Real Gravi-Particles:} In addition to
their virtual effects just mentioned, the gravi-particles can
decay into brane-matter or can be produced by scatterings among
the brane matter \cite{giudice,han,fr}. While the former plays a
crucial role in cosmological and astrophysical contexts, the
latter constitutes one of the most important signatures of extra
dimensions at colliders in that gravi-particle emission from the
brane gives rise to scattering processes with single missing
energy signal, and thus, it is of fundamental importance for
distinguishing supersymmetric models from the extra dimensional
ones. In these real gravi-particle involving processes the
presence of bulk masses for tensor and scalar modes modify decay
signatures significantly, as has been detailed in \cite{fr}.

For a clearer view of the effects of gravi-particle decay/emssion
it proves useful to refer to their loop effects. Indeed, the $Z$
boson self-energy, for example, represents, via the optical
theorem, the Drell-Yan production of gravi-particles and $Z$ boson
at lepton (via $e^+ e^- \rightarrow Z^{\star} \rightarrow
gravi-particle\; + Z$ annihilation) or hadron (via $q \overline{q}
\rightarrow Z^{\star} \rightarrow gravi-particle\; + Z$
annihilation) colliders. The main novelty brought about by
$f\left({\cal{R}},P,Q\right)$ gravity is the production of $J=2$
ghost and the scalar mode when the center-of-mass energy of the
collider is sufficiently large. This phenomenon reflects by itself
a sudden change in the number of events (similar to opening of
$W^+ W^-$ channel at LEP experiments). The dominant contribution
to gravi-particle emission comes from Kaluza-Klein levels in the
vicinity of $R^2 (M_Z^2 - m_{gravi-particle}^2)$. However, one
here notes an important aspect of gravi-particle decay/emission
processes: For such processes all gravi-particles must be fields
with positive semi-definite mass-squareds and hence, as in
$f\left({\cal{R}}\right)$ gravity \cite{fr}, one does not expect
significant contributions from gravi-particles other than $J=0$
graviton (see the figures Fig. \ref{boncuk_1}--\ref{boncuk4}).

\section{Conclusion}
In this work we have discussed phenomenological implications of
$f({\cal{R}},P,Q)$ gravity in higher dimensional spacetimes with
large extra spatial dimensions. In Sec. 2 we have expanded action
around a flat background and computed the propagator. Moreover,
after determining the propagating degrees of freedom and virtual
gravi-particle exchange amplitude we have provided a detailed and
comparative analysis of the contributions of $f({\cal{R}},P,Q)$
and Einstein gravity. We have therein witnessed important
enhancements/suppressions, as illustrated via the figures, brought
about by the higher-curvature gravity theory considered.

In Sec. 3 we have analyzed effects of virtual and real
gravi-particles on the scatterings among the brane matter. This
section has shown that there exist a number of laboratory and
astrophysical processes a global analysis of which can provide
important information about the nature of the gravitational theory
in the higher dimensional bulk. The discussions therein suggest
that $f({\cal{R}},P, Q)$ gravity theories with finite $f_{RR, P,
Q} (0)$ can induce potentially important effects testable at
future collider studies.

The higher-curvature gravity theory discussed in this work offer
various signatures which distinguish it from the Einstein and
$f({\cal{R}},P,Q)$ gravity theories, and a global survey of
laboratory, astrophysical and cosmological observables (see
\cite{large-extra} can reveal presence or absence of such
higher-curvature generalizations of the Einstein-Hilbert action.
Indeed, affirmative or negative, the answer will be crucial for
establishing the gravitational interactions beyond Fermi energies
and may pave the road to a full understanding of quantum gravity.

\section{Acknowledgements}
 The work of D. A. D. was partially supported by Turkish Academy of Sciences through GEBIP grant,
 and by the Scientific and Technical Research Council of Turkey through project 104T503.

\end{document}